\documentstyle[11pt]{article}

\begin{document}
\topmargin 0pt
\oddsidemargin 5mm
\setcounter{page}{1}
\begin{titlepage}
\hfill Preprint YeRPHI-1524(24)-98

\hfill October 29

\vspace{2cm}
\begin{center}

{\bf
Electroweak Lepton-Lepton and Lepton-Antilepton Bound States }\\
\vspace{5mm}
{\large R.A. Alanakyan}\\
\vspace{5mm}
{\em Theoretical Physics Department,
Yerevan Physics Institute,
Alikhanian Brothers St.2,

 Yerevan 375036, Armenia\\}
 {E-mail: alanak @ lx2.yerphi.am\\}
\end{center}

\vspace{5mm}
\centerline{{\bf{Abstract}}}
In model independent way we consider the possibility of the 
existence of fermion-antifermion,
fermion-fermion bound states which appear 
due to $\gamma,  Z^0(W^{\pm}$-bosons and scalar, pseudoscalars 
 exchanges including radiative corrections. Our consideration includes the case 
where at least one particle in the bound state is Majorana fermion or scalar.
This paper considers various types of bound states of particles contained in 
various extensions of the Glashow-Weinberg -Salam model.
We calculate long-range forces induced by photonic, gluonic and fermionic loop
corrections in scalar (pseudoscalar) exchange.

\vspace{5mm}
\vfill
\centerline{{\bf{Yerevan Physics Institute}}}
\centerline{{\bf{Yerevan 1998}}}

\end{titlepage}

                 {\bf 1.Introduction}

In this article by model independent way 
we consider possibility of existence of fermion-antifermion,
fermion-fermion bound states which appear due 
to $\gamma, Z^0(W^{\pm}$-bosons and scalar, pseudoscalars 
 exchanges including radiative corrections 
(long- range potential induced by neutrinos loop(i.e. Z(W)-exchanges including box diagrams) 
has been 
considered in \cite{FS}-\cite{SF}, 
long-range forces mediated by Higgs and Goldstown
 bosons exchange \cite{FN}(via loop of very light pseudoscalars):

We also consider (sect.4) a case  where at least
 one of the particles is Majorana neutrino (or neutralino)
 ( $\l^{\pm}_i \bar{\nu_j},
\l^{\pm}_i \bar{N_j},
 \nu_i\bar{\nu_j}
,\nu_i N_j,$,
$\tilde{\chi^0_i}\tilde{\chi^0_j}$,$\tilde{\chi^0_i}\tilde{\chi^{\pm}}$ ,
etc bound states) which contains in many extentions beyond 
the Standard Model  e.g. in the Minimal Supersymmetric Standard Model \cite{HK},\cite{GH}, 
or in Left-Right Symmetric Model \cite{PS}-\cite{M2}.

Our consideration ia also true for bound state of scalar neutrinos 
\footnote {The bound state $\tilde{\nu}\tilde{\nu^*}$ 
has been considered in \cite{KKT}, 
however in this paper the authors do not considered $Z^0$-exchanges and
 also do not consider radiative corrections to Higgs exchange which as 
we will see below radically change the Higgs potentials.}

As known, at tree level level $Z^0$ -bosons exchange lead to Yukawa potential :
\begin{equation}
\label{A1}
Z_0(r)=\frac{eg_V^c}{4\pi}\frac{\exp(-m_Zr)}{r}
\end{equation}
where 
$g_V^c=\frac{1}{s_Wc_W}(T_c-2Qs^2_W)$ (if attractive center is fermion), 
$s_W,c_W$- cosinus and sinus of 
Weinbergs angle, $T_c,Q_c$-are isospin and charge of attractive center.
In [1] it has been shown that at large distances the potential
 enhanced due to neutrino -pairs exchanges:
\begin{equation}
\label{A2}
V(r)=\frac{G_F^2}{4\pi^2r^5}
\end{equation}
Thus,  radiative corrections makes the potential 
deeper and wider and chances that bound state may exist are increasing.
We extend in this paper also the result [1] in range $r \sim \frac{1}{m_Z}$
or smaller $r$.

 As we will see below analogous behaviour takes place also for Higgs potential:
at small distances it is Yukawa-like, at large distances radiative corrections
 lead to slower decreasing $\sim r^{-5}$, and to behaviour $\sim r^{-7},r^{-9}$
for scalar (pseudoscalar) exchange for correction 
via photonic and fermionic loop. 

                                 {\bf 2.Potential}

As mentioned above at large distances  $r>> \frac{1}{m_Z}$ for potential
 between two electrons in accordance with [1] is described by (2).

At distances $r<\frac{1}{m_{\nu}}$ the contribution of neutrinos with masses 
$m_{\nu}$ is supressed as $\sim \exp(-2m_{\nu}r)$.

If we take into account not only neutrino [1] but all light fermions in loop
 we obtain at $\frac{1}{2m_f}<< r << \frac{1}{m_Z}$:
\begin{equation}
\label{A4}
Z_0(r)=\frac{3eg_V^c}{\pi^2r^5}\frac{\Gamma}{m^5_Z}
\end{equation}
where $\Gamma$ is the width of $Z^0$-boson.If in this formula we take into account only
neutrinos in loop and besides consider box and triangle diagramm we turn to formula (2).

At small distances $r<< \frac{1}{m_Z}$ from (5) as expected may be obtained 
well known result:
\begin{equation}
\label{A5}
V(r)=\frac{eg_V^c}{4\pi r}
(1+\frac{2}{\pi}\frac{\Gamma}{m_Z}\log(rm_Z))
\end{equation}
In fact, in (3),(4) all light fermions contributions are included.
It must be noted also that in formulas (3),(4) we take into account 
only $t$-channel diagram (Fig.1) 
in which all $N\approx\Gamma/\Gamma(Z^0\rightarrow\nu\nu)$
flavours contribute, whereas only one flavour contributes into the
box and triangle diagramm.
Both formulas (3),(4) may be obtained from the general expression 
(in contrast to the [1] we do not neglect $t$' in $Z^0$-bosons propagators):
\begin{equation}
\label{A6}
Z_0(r)=-\frac{eg^c_V}{4\pi^2r} 
\int \limits_{4m^2_f}^{\infty} dx \frac{ImP_T(t)\exp(-r\sqrt{t})}
{(t-m^2_Z)^2+(ImP_T(t))^2},
\end{equation}
where at $t>>4m_f^2$ for 
imaginary part of polarisation transversal operator we have:
\begin{equation}
\label{A7}
Im P_T= -\frac{\Gamma}{m_Z}t
\end{equation}
For instance, if  $r\gg\frac{1}{m_Z}$ for $t's$ in propagators may be neglected and
we obtain (3). At  $r\gg\frac{1}{m_Z}$ the main contribution in integral (5) comes 
from the range where
$r\sqrt{t}\ll 1$.Besides, at small distances we can neglect light fermion masses
$4m^2$ and integrate in (5) within the limits $0<t<r^{-2}$ and put 
$e^{-r\sqrt{t}}\approx 1$.
After the integration we obtain the result (4). 

At large distances $r\sim\frac{1}{m_f}$ we must
 take into account fermions masses in loop,
however, as in previous reference, again $q^2$ and $P_T$ 
in propagator may be neglected, and analogously [5] we obtain
\begin{equation}
\label{A8}
V(r)=\frac{3\alpha g^c_V}{2\pi^3r^2m_Z^5}\sum_f \frac{\Gamma(Z\rightarrow f\bar{f})}{\Gamma}K_2(2m_fr),
\end{equation}
where $K_2(x)$ is modified Bessel function.
At distances $r>\frac{1}{m_f}$ the contribution of the flavour $f$ falls as 
$\sim e^{-2rm_f}$.

Using the general formula (5) we can obtain also 
long-range forces mediated by virtual photons (virtual gluons) loops\cite{A}:

$H^0(P^0)\rightarrow \gamma^*\gamma^*\rightarrow H^0(P^0)$,

$H^0(P^0)\rightarrow gluon^*gluon^*\rightarrow H^0(P^0)$ ,

or by light fermions loop :

$H^0(P^0)\rightarrow \bar{f}^*f^*\rightarrow H^0(P^0)$.

This exchanges are shown on Fig.3

Substituting imaginary part of $H^0(P^0)$-bosons polarisation  operator:
\begin{equation}
\label{A9}
Im P_{H,P}= -\frac{\Gamma(H^0(P^0)\rightarrow\gamma\gamma(gg))
}{m^3_{H(P)}}t^2
\end{equation}
into (5) we obtain 
(for Higgs bnosons decays see e.g. \cite{O} and references therein, for $\pi^0$ decays see \cite{S},\cite{BJ}):
\begin{equation}
\label{A10}
V_{S,P}(r)=- a_{S(P)}(1)a_{S,P}(2)\frac{5!}{2\pi^2}
\frac{\Gamma(H^0(P^0)\rightarrow\gamma\gamma(gg))}{(m_{H(P)}r)^7}
\end{equation}
Thus, instead of  Yucawa potential from tree exchange we have long-range 
forces which decrease as 
$\sim r^{-7}$.
For arbitrary $r$ take place formula like (5).The potentials $V_{S,P}$
are contains into relativistic equations (see below).

In case of
 $\pi^0$-meson 
(which is in fact a composite 
pseudoscalar) formula true only at 
$r>m^{-1}_{\pi}$, at 
$r<m^{-1}_{\pi}$ the compositness of $\pi^0$-meson became considerable.

In non-relativistic approximation for $P^0$-bosons exchanges between fermions 
we have:
\begin{equation}
\label{A11}
V_P(r)=-a_P(1)a_P(2)\frac{5!}{2\pi^2}\frac{1}{4m_1m_2}
\frac{\Gamma(P^0\rightarrow\gamma\gamma(gg))}{(m_{P}^7r^9)}
(-7\vec{\sigma_1}\vec{\sigma_2}
+\frac{63(\vec{\sigma_1}\vec{r})(\vec{\sigma_2}\vec{r})}{r^2})
\end{equation}
where $\sigma_{1,2},m_{1,2}$ are operators of spins and masses of the fermions,
$a_{S,P}(1),a_{S,P}(2)$ its couplings with $P^0$-bosons.

Thus, we have long range potential which fall as $r^{-9}$.  

For fermion loop-induced Higgs potentials we obtain the following
model-independent result(at $r>>\frac{1}{2m_f}$):
\begin{equation}
\label{A12}
V_{S,P}(r)=-\frac{3a_{S,P}(1)a_{S,P}(2)}{\pi^2}
\frac{\Gamma(H^0(P^0)\rightarrow f\bar{f})}{m_{H(P)}^5r^5}
\end{equation}
Again ,at 
$r< \frac{1}{2m_f}$ 
the potentials (12) supressed as 
$\sim e^{-2rm_f}$.

In this form both potentials induced by fermionic loops are
 contains in equations (19)-(25) below.

In non-relativistic approximation however pseudoscalar 
exchange (with fermios loop) between fermions has the form:
\begin{equation}
\label{A13}
V_P(r)=-a_P(1)a_P(2)\frac{5!}{2\pi^2}\frac{1}{4m_1m_2}
\frac{\Gamma(P^0\rightarrow f\bar{f})}{(m_{P}^7r^7)}
(-5\vec{\sigma_1}\vec{\sigma_2}
+\frac{35(\vec{\sigma_1}\vec{r})(\vec{\sigma_2}\vec{r})}{r^2})
\end{equation}
while the Higgs potential in non-relativistic approximation again defined by 
formula (9).

In \cite{A}  is given comparision of long-range forces (10) 
induced by $\pi^0$-meson exchange with various 
other kinds of long-range forces including Van-Der-Vaals potential.

The presented above consideration for scalar and pseudoscalars exchanges 
is model independent and applicable also to potentials created by scalars
leptons exchanges ((s)fermion -(s)fermion-scalar lepton vertexes arising 
in theories with $R$-parity violation (see ref.\cite{F}-\cite{D} ).
In this case $a_{S,P}$ may be not proportional to fermion masses.

{3.\bf Dirac Particle Case}

The Dirac equation\footnote{for Dirac equation in 
$Z^0$-boson field for Dirac and Majorana spinors 
see e.g. \cite{P1} and references therein} for 
particle in spherically symmetric potentials 
$Z_0,A_0,V_{S,P}$ has the form:
\begin{equation}
\label{A18}
(\hat{k}-e\hat{Z_0}(g_V+g_A\gamma_5)-eQ\hat{A}-m+a_SV_S+a_P\gamma_5fV_P )u(k)=0,
\end{equation}
Here $Z_0(r)A_0(r),V_{S,P}(r)$ are potentials created 
by immovable center  and described in sect.2,
 $g_V=\frac{1}{c_Ws_W}(T-2Qs^2_W),g_A=\frac{1}{c_Ws_W}2T$, 
where $T,Q$ are isospin and charge of particle respectively.

In (13) $V_{S,P}$ defined in sect.2 above, $a_{S,P}$ are model independent 
Yukawa couplings of scalars and pseudoscalars with fermions.
In $a_{S,P}$ we include also scalars (pseudoscalars) interaction with
attractive center.

If we consider a moving antiparticle instead of a particle we must 
make the replacements
in above Dirac equation :
\begin{equation}
\label{A19}
g_V,g_A,Q \rightarrow -g_V,-g_A,-Q
\end{equation}
Let us consider the movement of fermion (antifermion) in spherically symmetric potentials $A_0,Z_0,V_{S,P}$
using the method developed in\cite{AB} \cite{LL4}(see also references therein).

Because electroweak interactions violated $P$-parity (term$\gamma_{\mu}\gamma_5$
in Dirac equation), spinors $\phi,\chi$ must be expressed through linear combination of spheric spinors
$\Omega_{jlM}(\vec{n}),\Omega_{jl'M}(\vec{n})$ which have different $P$-parity:
\begin{eqnarray}
\label{A20}
&&\psi^T=
(\phi,\chi),\nonumber\\&& 
\phi=f_1(r)\Omega_{jlM}(\vec{n})+
(-1)^{\frac{1+l-l'}{2}}
f_2(r)\Omega_{jl'M}(\vec{n}),\nonumber\\&&
\chi=g_1(r)\Omega_{jlM}(\vec{n})+
(-1)^{\frac{1+l-l'}{2}}
g_2(r)\Omega_{jl'M}(\vec{n}))
\end{eqnarray}
where $l=j\pm \frac{1}{2}$, $l'=2j-l$.
Using identies:
\begin{equation}
\label{A21}
\vec{\sigma}\vec{p}\phi=
(g_1'(r)+\frac{1+\kappa}{r}g_1(r))\Omega_{jl'M}(\vec{n})-
(g_2'(r)+\frac{1-\kappa}{r}g_2(r))\Omega_{jlM}(\vec{n})
\end{equation}
\begin{equation}
\label{A22}
\vec{\sigma}\vec{p}\chi=
(f_1'(r)+\frac{1+\kappa}{r}f_1(r))\Omega_{jl'M}(\vec{n})-
(f_2'(r)+\frac{1-\kappa}{r}f_2(r))\Omega_{jlM}(\vec{n}),
\end{equation}
where
\begin{equation}
\label{A23}
\kappa=l(l+1)-j(j+1)-\frac{1}{4},
\end{equation}
for radial functions we obtain:
\begin{equation}
\label{A24}
(g_2'(r)+\frac{1-\kappa}{r}g_2(r))+(E-M(r)-V(r))f_1(r)+V_-(r)g_1(r)=0
\end{equation}
\begin{equation}
\label{A25}
(g_1'(r)+\frac{1+\kappa}{r}g_1(r))
-(E-M(r)-V(r))f_2(r)+V_-(r)g_2(r)=0
\end{equation}
\begin{equation}
\label{A26}
(f_2'(r)+\frac{1-\kappa}{r}f_2(r))+(E+M(r)-V(r))g_1(r)+V_+(r)f_1(r)=0
\end{equation}
\begin{equation}
\label{A27}
(f_1'(r)+\frac{1+\kappa}{r}f_1(r))-(E+M(r)-V(r))g_2(r)-V_+(r)f_2(r)=0
\end{equation}
\begin{equation}
\label{A28}
M(r)=m-a_SV_S(r)
\end{equation}
\begin{equation}
\label{A29}
V(r)=eg_VZ_0(r)+eQA_0(r),
\end{equation}
\begin{equation}
\label{A30}
V_{\pm}(r)=eg_AZ_0(r)\pm a_PV_P(r),
\end{equation}
For large $r$ we can drop all fields
 and obtain divergent spheric waves at infinity.

Although above we consider stationar solutions, our consideration is also
applicable for time-dependent solutions.In this case we must in (19)-(22)
make the replacements:
\begin{equation}
\label{A31}
E\rightarrow -i\frac{d}{dt},
\end{equation}
and 
\begin{equation}
\label{A32}
f_i(r)\rightarrow f_i(r,t),
\end{equation}
\begin{equation}
\label{A33}
g_i(r)\rightarrow g_i(r,t),
\end{equation}
At small $r\ll \frac{1}{m_Z}$ we can put 
$Z_0(r)=\frac{eg_V^c}{4\pi r}$,
$V_{S,P}(r)=\frac{1}{r}, E=0$.
For radial wave functions we suppose that the solutions have the following form:

Substituting (29) into (19)-(25) we obtain:
\begin{equation}
\label{A34}
g_i(r)=a_i r^{\gamma},
f_i(r)=b_i r^{\gamma},
\end{equation}
\begin{equation}
\label{A35}
(\gamma+1-\kappa)a_2+(a_S-\alpha(g_Vg_V^c+QQ_c)b_1+a_1(\alpha g_Ag_V^c-a_P)=0
\end{equation}
\begin{equation}
\label{A36}
(\gamma+1+\kappa)a_2-(a_S-\alpha(g_Vg_V^c+QQ_c)b_2+a_2(\alpha g_Ag_V^c-a_P)=0
\end{equation}
\begin{equation}
\label{A37}
(\gamma+1-\kappa)b_2+(-a_S-\alpha(g_Vg_V^c+QQ_c)a_1+b_1(\alpha g_Ag_V^c+a_P)=0
\end{equation}
\begin{equation}
\label{A38}
(\gamma+1+\kappa)b_1+(a_S+\alpha(g_Vg_V^c+QQ_c)a_2-b_2(\alpha g_Ag_V^c+a_P)=0
\end{equation}
The determinant of this system of equations  
must be equal zero, and that condition
 gives us the equation for defining $\gamma$.

The general solution of this equation is very complicated. 
For simplicity consider several cases:

1)$a_{S,P}=Q=0$.

2)Only $a_S$ is nonzero.

3)only $a_P$ is nonzero.

In case of 1) we obtain 4 solutions:
\begin{equation}
\label{A39}
s=-1+\sqrt{\kappa^2-\alpha^2(g_V^2\pm g_A^2)}
\end{equation}
\begin{equation}
\label{A40}
s=-1-\sqrt{\kappa^2-\alpha^2(g_V^2\pm g_A^2)}
\end{equation}
In case of 2),3) the full system of equations decouples on two 
subsystems of equations with solutions:
\begin{equation}
\label{A41}
s=-1\pm\sqrt{\kappa^2-a_{S,P}^2}
\end{equation}
\begin{equation}
\label{A42}
s=-1\pm\sqrt{\kappa^2+a_{S,P}^2}
\end{equation}
In case of pure QED we obtain again 
$s=-1\pm\sqrt{\kappa^2-Q^2Q_c^2\alpha^2}$.

Obviously, if $\alpha\ll 1$, the solutions (35) must be excluded because at 
near $r=0$ the radial functions behaviour are 
$R(r)\sim r^{-1-|\kappa|+O(\alpha^2)}$.Analogously by the same reasons
 $s=-1-\sqrt{\kappa^2\pm a_{S,P}^2}$ must be excluded if $a_{S,P} \ll 1$.
The solution 
$s=-1-\sqrt{\kappa^2+a_{S,P}^2}$ 
must be excluded for any $a_{S,P}$.

{\bf 4. Majorana Particle Case}

For Majorana particle case we must put in equation (13) and equations 
(19)-(26) $g_V=Q=0$:
\begin{equation}
\label{A43}
(\hat{k}-g\hat{Z_0}g_A\gamma_5-m+a_SV_S+a_P\gamma_5fV_P )\psi_M(k)=0,
\end{equation}
whith Majorana spinors:
\begin{equation}
\label{44}
\psi_M=\frac{1}{\sqrt{2}}(\psi+\psi^c)
\end{equation}
where $\phi,\chi$ defined above in (15).

In case of right-handed heavy neutrino 
$Z^0NN,W^+l^-\nu$ vertexes 
are supressesed by the smallness of light neutrino 
mass and therefore heavy neutrinos bound state exist predominantly 
due to Higgs bosons
 exchange ($Z^0_R$-boson interact with heavy neutrino, however if it is 
heavier than Higgs bosons, its contribution is negligible).
Also , due to the smallness of 
$Z^0N\nu,W^+l^-\nu$ vertexes widths of the decays
$N\rightarrow Z^0_L\nu,W^{\pm}l^{\mp}$ 
are small in comparision with energy levels.

                 {\bf 5.Scalars}

Our consideration is also applicable to scalar neutrinos 
bound state which appear e.g.in Minimal Supersymetric Standard Model 
(for  Minimal Supersymetric Standard Model 
see \cite{HK},\cite{GH} and references therein)
$\tilde{ \nu}_i\bar{\tilde{\nu}_j},
\tilde{ \nu}_i\bar{\tilde{\nu}^*_j},
\tilde{\chi}^0_i\tilde{\chi}^0_j$
bound states of scalar+ pseudoscalar.

Klein-Gordon equation of motion for scalar neutrino (antineutrino)
 in potentials 
$Z_0,V_{S,P}$  takes the form: 
\begin{equation}
\label{A50}
((E-e(g_VZ_0))^2-\vec{k}^2-m^2+b_SV_S+b_PV_P)u(k)=0,
\end{equation}
At small $r$ we find the solution in the form:
\begin{equation}
\label{A54}
g_i(r))=a_ir^s,
\end{equation}
where only the most non-singular solution must be taken into account:
\begin{equation}
\label{A55}
s=\frac{1}{2}(-1+\sqrt{1-\alpha^2(g_Vg_V^c)^2})
\end{equation}
Depending on sign of $g_V$, the equation (40) describes scalar 
neutrino-scalar neutrino, scalar neutrino-scalar antineutrino systems.

Now we make numerical calculation which will be described in details 
in our next paper \cite{AP}.

                 {\bf 7. Bethe-Salpeter equations}

If $Q_a,Q_b,T_a,T_b $ are charges and isospins of two Dirac
particles a and b, the Bethe-Salpeter (see e.g.\cite{W} and references therein, 
for Bethe Salpeter equation for scalars systems) 

equation of the such system has the following form:

\begin{eqnarray}
\label{A70}
&&(\hat{k_a}-m_a)\chi(\hat{k_b}-m_b)=e^2
\int \limits \frac{d^4k}{(2\pi)^4} \frac{(g^a_V\gamma_n+g^a_A\gamma_n\gamma_5) \chi(\gamma_ng_V^b+g_A^b \gamma_n\gamma_5)}
{(\vec{p}-\vec{k})^2+M^2+P_T}
\nonumber\\&& 
+e^2Q_aQ_b
\int \limits \frac{d^4k}{(2\pi)^4} \frac{\gamma_n \chi \gamma_n}
{(\vec{p}-\vec{k})^2}+
\sum a_P^aa_P^b\int \limits \frac{d^4k}{(2\pi)^4} \frac{\gamma_5 \chi \gamma_5}{(\vec{p}-\vec{k})^2+m_P^2+P_P}
+\nonumber\\&& 
\sum a_S^aa_S^b\int \limits \frac{d^4k}{(2\pi)^4} \frac{\chi}{(\vec{p}-\vec{k})^2+m_P^2+P_S},
\end{eqnarray}
where 
$g_V^{a,b}=\frac{1}{c_Ws_W}(T^{a,b}_3-2Qs^2_W),
g_A^{a,b}=\frac{1}{c_Ws_W}2T^{a,b}_3$.
    
If one of the particle is Majorana neutrino we must put in the above
equation: 
\begin{equation}
\label{A71}
g^a_V=Q_a=0,
\end{equation}
if a is  Majorana neutrino;
\begin{equation}
\label{A72}
g^b_V=Q_b=0,
\end{equation}
if b is  Majorana neutrino; and
\begin{equation}
\label{A73}
g^{a,b}_V=Q_{a,b}=0,
\end{equation}
if both particles are Majorana fermions.

For the same flavour particles bound states (e.g. like
$l^-_i,l^-_i$) we must take into account also $u,s$-channels
exchanges (see Fig.1).It must be noted that 
other loops with triangle vertex and four
-point integrals (Fig.1) which are of order
$O(\frac{G_F^2}{r^5})$ at large ($r \gg \frac{1}{m_Z}$) distances,
should be taken into account too.

Analogously, for scalar patricles case we obtain:
\begin{eqnarray}
\label{A74}
&&(\hat{k_a^2}-m_a^2)\chi(\hat{k_b^2}-m^2_b)=g_V^ag_V^b
\int \limits \frac{d^4k}{(2\pi)^4} \frac{(k_a+q)(2k_b+k_a-q)\chi}
{(\vec{q}-\vec{k_a})^2+m_Z^2+P_T}
\nonumber\\&& 
+e^2Q_aQ_b
\int \limits \frac{d^4k}{(2\pi)^4} \frac{\chi}
{(\vec{p}-\vec{k})^2}+
\nonumber\\&& 
\sum b^a_Sb^b_S\int \limits \frac{d^4k}{(2\pi)^4} \frac{\chi}{(\vec{p}-\vec{k})^2+m_P^2+P_S},
\end{eqnarray}
where coefficients $b_S$- are Yukawa couplings of scalar leptons
 to Higgs bosons (see e.g. \cite{HK} 
for Higgs bosons- scalar leptons interaction within MSSM).

In theories with R-parity violation it is possible also 
$\tilde{\nu}\tilde{l}\tilde{l}$ couplings are also possible.
  In Bethe-Salpeter equation for fermions in this case we obtain terms 
$\chi\gamma_5$ or$\chi\gamma_5$  which lead to $P$-parity violation
which will manifested in nonrelativistic lagrangian 
as $V(r) \sim \sigma \vec{r}$ \cite{A}.

                 {\bf Acknowledgements}

The authors express his sincere gratitude to all participants
 of the seminar  which took place in Yerevan Physics Institute on 
October 29, 1998 and to I.Aznauryan, A.Allakhverdyan, G.Griroryan, 
A.Melikyan, 
R.Pogossyan, D.Sahakyan, G.Yegiyan, E.Prokhorenko for 
fruitful discussions.

\end{document}